\documentclass[aps,preprint]{revtex4}%
\usepackage{amsfonts}
\usepackage{amsmath}
\usepackage{amssymb}
\usepackage{graphicx}%
\begin{document}
\preprint{ }
\title{Vortex-glass transition in superconducting Nb/Cu superlattices}
\author{J.E. Villegas}
\affiliation{Dpto. F\'{\i}sica de Materiales, Universidad Complutense de Madrid, 28040
Madrid, Spain}
\author{J.L. Vicent}
\affiliation{Dpto. F\'{\i}sica de Materiales, Universidad Complutense de Madrid, 28040
Madrid, Spain}

\begin{abstract}
Nb/Cu superconducting superlattices have been fabricated by dc magnetron
sputtering. This system shows a vortex glass transition with critical
exponents similar to high temperatures superconductors exponents. The
transition dymensionality is governed by the superconducting coupling regime.
The vortex glass transition shows a \textit{pure} two dimensional behavior in
decoupled superlattices and a \textit{quasi}-two dimensional behavior in the
superlattice coupling regime.

\end{abstract}
\maketitle

Since the discovery of High-Tc Superconductors (HTCS) vortex matter Physics
has called the attention of researchers in many fields, for example the vortex
state provides a perfect realm to study properties of liquid, crystalline and
glassy phases. A plethora of different phases is observed in HTCS,$^{1}$
induced by the interplay of vortex-vortex interaction, thermal fluctuations,
different kinds of disorder, anisotropy, dimensional effects, etc.$^{2}$ But
the most remarkable characteristic of the phase diagram is the existence of
two different states: a magnetically irreversible zero-resistance state, and a
reversible state with dissipative transport properties. In the absence of
disorder, the former corresponds to a vortex-solid phase with topological
order, and the latter to a vortex-liquid phase, both of them separated by a
first-order melting transition. In the presence of strong quenched disorder,
however, the topological order of the vortex-lattice is lost, and the
zero-resistance state corresponds to a vortex-glass (VG), and the transition
into the dissipative liquid state becomes a continuous second-order phase
transition .$^{3,4}$

After the rich phenomenology of vortex-phases came out with HTSC, Low-Tc
Superconductors (LTCS) have been seldom revisited to check the possible
application of \ HTCS paradigms. The existence of a melting transition in Nb
single crystals has been reported,$^{5-7}$ \ but in the case of Nb thin films
the existence of a glass transition remains controversial.$^{8,9}$

In this paper, we report on the observation of VG transitions in sputtered
Nb/Cu low-temperature superconducting superlattices. This system has been
chosen mainly because the presence of quenched disorder is similar to the one
existing in sputtered Nb thin films and because the artificially layered
structure allows tailoring the anisotropy of the system in a controllable
fashion. In particular, the coupling of Nb layers through the Cu spacer can be
easily tuned.$^{10}$ This has allowed us studying the dimensionality of the VG
transition in several different coupling regimes.

The VG transition has been investigated by measuring the electrical transport
properties in the mixed state of superlattices, in particular isothermal I-V
characteristics in applied magnetic fields. These characteristics have been
collapsed, according to the scaling rules proposed in the VG theory, in terms
of critical exponents.$^{4}$ Besides, the consistency of the scaling analysis
has been checked with some independent criterion, as recently proposed.$^{11}$
Moreover, we have found a dimensional crossover for a \textit{quasi}-2D into a
\textit{pure} 2D VG transition, governed by the coupling of Nb layers in the superlattice.

Nb/Cu superlattices were grown on Si (100) substrates using \textit{dc}
magnetron sputtering at room temperature in Ar atmosphere. Several series of
superlattices Cu$_{\mathrm{d}}$[Nb$_{\mathrm{d}\mathrm{^{\prime}}}%
$/Cu$_{\mathrm{d}}$]$_{\mathrm{N}}$ were grown with N being the number of
bilayers. Structural characterization was made by X-Ray diffraction (XRD) .
The structural properties of Nb/Cu superlattices have been investigated early
by other authors.$^{12}$ In our multilayers XRD shows that Cu layers are
oriented (111), while Nb ones are (110). We have refined the spectra using
SUPREX program.$^{13}$ From refinements the modulation length $\Lambda$, as
well as several sources of disorder at the interfaces, like roughness or
interdiffusion have been obtained. The samples studied here did not present
interdiffusion, and have moderate roughness at the interfaces, ranging from
0.2 to 0.6 nm, being larger when the Nb layers are thicker. The samples were
lithographed by wet etching into a measuring bridge 1 mm long and 100 $\mu$m
wide for magnetotransport experiments with standard four-probe configuration.

Magnetotransport experiments were made in a liquid He cryostat provided with a
superconducting solenoid. The superconducting coherence length \textit{$\xi$%
}$_{S}$ as a function of temperature was calculated from measured upper
critical fields $H_{c2}(T)$, obtained from magnetoresistance measurements at
constant fixed temperatures $R(H)_{T}$. The in-plane coherence length
\textit{$\xi$}$_{S||}$ (parallel to Nb/Cu interfaces) and the perpendicular
one \textit{$\xi$}$_{S\bot}$ have been calculated by using $\xi_{S||}\left(
{T}\right)  ={\left[  {{{\phi_{0}}\mathord{\left/ {\vphantom {{\phi _{0}}
{2\pi H_{c2 \bot}  \left( {T} \right)}}} \right. \kern-\nulldelimiterspace}
{2\pi H_{c2 \bot} \left(  {T} \right)  }}}\right]  }^{1/2}$ and $\xi_{S\bot
}\left(  {T}\right)  ={\left[  {{{\phi_{0}H_{c2\bot}\left(  {T}\right)
}\mathord{\left/ {\vphantom {{\phi _{0} H_{c2 \bot}  \left( {T} \right)}
{2\pi \left( {H_{c2\vert \vert}  \left( {T} \right)} \right)^{2}}}} \right.
\kern-\nulldelimiterspace} {2\pi \left( {H_{c2\vert \vert}  \left( {T}
\right)} \right)^{2}}}}\right]  }^{{{1}\mathord{\left/ {\vphantom {{1} {2}}}
\right. \kern-\nulldelimiterspace} {2}}}$\textbf{\ }respectively.$^{10}$ As an
example, the observed behavior for sample Nb$_{\mathrm{1}\mathrm{3}%
\mathrm{n}\mathrm{m}}$/Cu$_{\mathrm{2}\mathrm{7}\mathrm{n}\mathrm{m}}$ is
shown in Fig. 1. The perpendicular critical field displays typical linear
dependence on temperature $H_{c2\bot} $\textit{(T)$\propto$(1-T/T}$_{c})$ (see
Fig. 1 inset). However, the parallel critical field shows up a crossover from
linear dependence $H_{c2||} $\textit{(T)$\propto$(1-T/T}$_{c})$ at high enough
temperatures to square-root $H_{c2||} $\textit{(T)$\propto$(1-T/T}$_{c}%
)^{1/2}$ at lower temperatures. A dimensional crossover takes place when the
perpendicular coherence length \textit{$\xi$}$_{S\bot}(T)$ reaches a value of
the order of $d_{\mathrm{C}\mathrm{u}}$ (thickness of Cu in the superlattice).
Below the crossover temperature, $T_{2D}\approx$0.5$T_{c}$, the Nb layers are
decoupled, showing up two-dimensional (2D) behavior.$^{10}$ In other cases, as
for instance Nb$_{\mathrm{3}\mathrm{.}\mathrm{4}\mathrm{n}\mathrm{m}}%
$/Cu$_{\mathrm{2}\mathrm{.}\mathrm{4}\mathrm{n}\mathrm{m}\mathrm{.}\mathrm{,}%
}$ \textit{$\xi$}$_{S\bot}(T)>d_{Cu}$ at all temperatures, and thus
superlattices are always in the coupled regime.

We have measured I-V characteristics with magnetic field $H$ applied
perpendicular to Nb/Cu layers. For each fixed value of applied field $H$, we
measured a set ($\sim$20) of isothermal I-V curves at decreasing temperatures,
as those shown in Fig. 2 and Fig. 3. For all measured samples and applied
magnetic fields the results are similar. The isotherm at the highest
temperature displays linear behavior at all current levels, with Ohmic
resistance $R=V/I\approx R_{n}$, at this temperature $H=H_{c2\bot}.$ For
characteristics at slightly lower temperatures, the Ohmic response is observed
only up to a threshold current level $I_{nl}$, above which the curves are
non-linear. The Ohmic resistance in the low-current limit
${\mathop {lim}\limits_{I\rightarrow0}}{\kern1pt}\;{{V}%
\mathord{\left/ {\vphantom {{V} {I}}} \right.
\kern-\nulldelimiterspace} {I}}\neq0$ is smaller as the temperature decreases,
as well as the onset of non-linear response $I_{nl}$ shifts to lower current
levels as temperature is reduced. Below a given temperature, isotherms become
highly non-linear within the whole experimental window, and isotherms show up
negative curvature in the low-current limit yielding zero resistance
${\mathop {lim}\limits_{I\rightarrow0}}{\kern1pt}\;{{V}%
\mathord{\left/ {\vphantom {{V} {I}}} \right. \kern-\nulldelimiterspace} {I}}%
=0$.

The phenomenology described above suggests the existence of a continuous
transition from a truly superconducting phase with zero-resistance (VG) to a
vortex-liquid dissipative phase closer to $H_{c2}$. As proposed by
Fisher-Fisher-Huse,$^{4}$ and shown experimentally for many HCTS systems, this
glass-transition is a second order transition and the physical quantities must
scale with the VG correlation length \textit{$\xi$}$_{VG}$ and the
characteristic relaxation time \textit{$\tau$}. These two magnitudes diverge
as temperature approaches the glass transition temperature $T_{g}$, following
$\xi_{VG}\propto\left(  {T-T_{g}}\right)  ^{-\nu}$ and $\tau\propto\left(
{T-T_{g}}\right)  ^{-z\nu}$, with $z$ and \textit{$\nu$} the dynamic and
static critical exponents. Scaling laws have been proposed to collapse onto a
single master curve all I-V (or $E-J$) isotherms within the critical region,
by means of the relation$^{4}$

$E\xi_{VG}\tau\approx J\xi_{VG}^{D-1}\zeta_{\pm}\left(  {{{J\phi_{0}\xi
_{VG}^{D-1}}\mathord{\left/ {\vphantom {{J\phi _{0} \xi
_{VG}^{D - 1}}  {k_{B} T}}} \right. \kern-\nulldelimiterspace} {k_{B} T}}%
}\right)  $ [Eq. 1]

\noindent where D is the dimensionality of the glass transition,
\textit{$\zeta$}$_{\pm}$ is a universal scaling function above (\textit{$\zeta
$}$_{+}$ ) or below (\textit{$\zeta$}$_{\mathrm{-}}$ ) $T_{g}$, \textit{$\phi
$}$_{0}$ the flux quantum, and $k_{B}$ the Boltzmann constant. We have applied
this scaling analysis to I-V characteristics measured within the experimental
window 10$^{\mathrm{-}\mathrm{8}}$V%
%TCIMACRO{\TEXTsymbol{<}}%
%BeginExpansion
$<$%
%EndExpansion
V%
%TCIMACRO{\TEXTsymbol{<}}%
%BeginExpansion
$<$%
%EndExpansion
10$^{\mathrm{-}\mathrm{4}}$V and I%
%TCIMACRO{\TEXTsymbol{<}}%
%BeginExpansion
$<$%
%EndExpansion
10$^{\mathrm{-}\mathrm{3}}$ A. The voltage cut-off is 10$^{\mathrm{-}%
\mathrm{4}}$ V, since above this limit all characteristics deviate towards
Ohmic behavior. Therefore, the scaling analysis is not longer
valid.$^{11,14,15}$ This can be seen in the inset of Fig. 2 (b), where the
slopes of isotherms [as plotted in Fig 2 (a)], ${{d\left(  {logV}\right)
}\mathord{\left/ {\vphantom {{d\left( {logV} \right)} {d\left( {logI}
\right)}}} \right. \kern-\nulldelimiterspace} {d\left( {logI} \right)}}$, are displayed.

For Nb/Cu superlattices whose Nb layers are coupled, the scaling analysis
yields very good collapses, as the one shown in Fig. 2 (b). Critical exponents
were around z $\approx$ 4 and $\nu\approx$1.8, and the dimensionality was
always D=2. These parameters are not magnetic field or sample dependent. The
values of the critical exponents z and \textit{$\nu$} are within the range 4-6
predicted by theory.$^{4}$ Moreover, we want to underline that the obtained
values are very similar to that observed for \textit{quasi}-2D VG transitions
in HTCS.$^{16,17}$ For instance, the same values of the critical exponents
were found in YBa$_{\mathrm{2}}$Cu$_{\mathrm{3}}$O$_{\mathrm{7}\mathrm{-}%
\mathrm{\delta}}$ thin films.$^{17}$ This fact stresses the universality of
the VG transition, earlier suggested.$^{17,18}$

Recently, Strachan et al.$^{11}$ have proved that the scaling method used is
misleading, even in the case of universal critical exponents with adequate
values were found. They have argued that experimental limitations related to
voltage sensitivity floor might allow achieving good collapses with arbitrary
values of some of the scaling parameters, as for instance $T_{g}$. Thus,
Strachan \textit{et al.}$^{11}$ have proposed that a new criterion to
unambiguously determine $T_{g}$ should be met: isotherms above and below
$T_{g}$, but with equal ${\left|  {{{\left(  {T-T_{g}}\right)  }%
\mathord{\left/ {\vphantom {{\left( {T - T_{g}}  \right)} {T_{g}} }} \right.
\kern-\nulldelimiterspace} {T_{g}}}}\right|  }$, must have opposite
concavities at the same applied current level. In our Nb/Cu superlattices this
criterion was checked for every set of I-V curves that had been successfully
scaled. In the inset of Fig 2 (b), the derivatives ${{d\left(  {logV}\right)
}\mathord{\left/ {\vphantom {{d\left( {logV} \right)} {d\left(
{logI} \right)}}} \right. \kern-\nulldelimiterspace} {d\left( {logI}
\right)}}$ of the I-V isotherms shown in Fig. 2 (b) are plotted. From the
scaling procedure we obtained T$_{\mathrm{g}}$=3.994 K for this set of
isotherms. As can be seen in that Figure, the isotherms above and below this
temperature (marked with an arrow in Fig. 2(d)) met the criterion. At the
lowest current levels, upward and downward isotherms are observed at similar
distance ${\left|  {{{\left(  {T-T_{g}}\right)  }%
\mathord{\left/ {\vphantom {{\left( {T -
T_{g}}  \right)} {T_{g}} }} \right. \kern-\nulldelimiterspace} {T_{g}}}%
}\right|  }$ below and above $T_{g}$. Furthermore the VG transitions theory
tells that the isotherm at $T_{g}$ fulfills the relation $E\propto
J^{\alpha+1}$ in the low current limit, where ${{\alpha=\left(  {z+2-D}%
\right)  }\mathord{\left/ {\vphantom {{\alpha = \left( {z + 2 - D}
\right)} {\left( {D - 1} \right)}}} \right. \kern-\nulldelimiterspace}
{\left(  {D - 1} \right)  }}$ .$^{4}$ Therefore, using the values obtained in
our scaling, $D$=2 and $z\approx$4, the expected slope of the critical
isotherm would be $\alpha+1\approx5$. Although the critical isotherm is not in
the set of measured characteristics, we can give lower and upper limits to its
slope from those of the isotherms just above and just below $T_{g}$. In the
low current limit (Fig. 2 (b) inset), the first isotherm below $T_{g}$ has a
slope larger than $\sim$5, whilst the first one above $T_{g}$ has a slope
smaller than $\sim$4.5 . Thus the critical isotherm in between them should
have a slope around $\sim$4.5-5, as expected from the independently obtained
scaling parameters.

The behavior of superlattices whose Nb layers were decoupled was investigated
too. The inset of Fig. 3 (a) shows I-V characteristics of sample
Nb$_{\mathrm{1}\mathrm{3}\mathrm{n}\mathrm{m}}$/Cu$_{\mathrm{2}\mathrm{7}%
\mathrm{n}\mathrm{m}} $ in applied magnetic field \textit{$\mu$}$_{0}H$=0.4 T,
such that all isotherms, within the critical region, were at temperatures well
below T=T$_{\mathrm{2}\mathrm{D}}\approx$0.5T$_{\mathrm{c}}$ (see Fig. 1).
Attempts to use the scaling rule (Eq. 1) did not yield collapses as the ones
obtained in the coupled regime. In fact, from the derivatives of I-V curves
[Fig. 3 (b)], one can observe some features that rule out the scaling of these
isotherms according to a \textit{quasi}-2D or 3D VG transition. On one hand,
it is not possible determining a finite \textit{Tg} using the criterion
outlined above, since two isotherms of opposite concavities at the same
current level can not be found. Besides, as can be seen in Fig. 3 (b), a
maximun slope a maximum slope of \ $\sim$9 is displayed by the lowest
temperature isotherm among those showing up a decreasing slope in the low
current limit (marked with an arrow). The isotherm at $T_{g}$ should be above
this one, and therefore it should have a slope larger that $\sim$9. As we said
before, in a \textit{quasi}-2D or 3D VG transition the slope of the critical
isotherm is ${{\alpha+1=\left(  {z+2-D}\right)  }%
\mathord{\left/ {\vphantom {{\alpha + 1 = \left( {z + 2 - D}
\right)} {\left( {D - 1} \right)}}} \right. \kern-\nulldelimiterspace}
{\left(  {D - 1} \right)  }}+1$, which would imply that the critical exponent
$z $%
%TCIMACRO{\TEXTsymbol{>}}%
%BeginExpansion
$>$%
%EndExpansion
8 for D=2, or $z$%
%TCIMACRO{\TEXTsymbol{>}}%
%BeginExpansion
$>$%
%EndExpansion
17 if D=3. These are very high values of the critical exponents, not supported
by theory.$^{4}$ As argued earlier,$^{16,17}$ this strongly suggests that a
\textit{quasi}-2D or 3D VG transition has to be dismissed. However, a good
scaling of the isotherms is achieved by assuming a different \textit{pure} 2D
VG transition, as first proposed by Dekker \textit{et al.}$^{19}$, and later
found in HTSC systems.$^{16,17}$ In a \textit{pure} 2D VG transition, the
glass transition temperature $T_{g}$=0. Therefore a \textit{pure} 2D VG phase
does not exist at any finite temperature, although in-plane correlations (2D)
develop, diverging as $\xi_{VG}\propto{{1}%
\mathord{\left/ {\vphantom {{1} {T^{\nu '}}}} \right.
\kern-\nulldelimiterspace} {T^{\nu '}}}$ when temperature approaches $T_{g}=0
$. In this transition, the scaling of the isotherms is achieved by plotting
$\rho exp{\left[  {\left(  {{{T_{0}}\mathord{\left/ {\vphantom
{{T_{0}}  {T}}} \right. \kern-\nulldelimiterspace} {T}}}\right)  ^{p}}\right]
}$ vs. ${{J}\mathord{\left/ {\vphantom {{J} {T^{1 + \nu '}}}}
\right. \kern-\nulldelimiterspace} {T^{1 + \nu '}}}$, where $\rho
={{E}\mathord{\left/ {\vphantom {{E} {J}}} \right. \kern-\nulldelimiterspace}
{J}}$ is the resistivity, $T_{\mathrm{0}}$ is a characteristic temperature,
and $p$ and \textit{$\nu$}'=2 are characteristic exponents of the
\textit{pure} 2D VG transition. The exponent $p$ is related to the mechanism
of vortex motion: $p\geq$1 for thermal activation over the relevant energy
barriers, whereas $p\approx$0.7 is expected in the case of quantum tunneling
across them.$^{19} $ As can be seen in Fig. 3 (a), a good collapse has been
obtained with parameters $T_{0}$=300$\pm$20 K, $p$=1.05$\pm$0.02, and
\textit{$\nu$}'=2.

Finally, in our experimental situation, a layered superconductor with magnetic
field applied perpendicular to layers, one may distinguish between the
in-plane VG correlation length \textit{$\xi$}$_{VG\mathrm{|}\mathrm{|}}$, and
the perpendicular one \textit{$\xi$}$_{VG\mathrm{\bot}}$(along the vortex
line).$^{4}$ The \textit{quasi}-2D character of the glass transition was
explained in high T$_{\mathrm{c}}$ superconductors assuming that anisotropy
induces limited vortex length, that precludes \textit{$\xi$}$_{VG\mathrm{\bot
}}$ to diverge .$^{16,17}$ Thus, when approaching $T_{g}$, only \ \textit{$\xi
$}$_{VG\mathrm{|}\mathrm{|}}$ diverges up to the macroscopic size of the
sample, whereas \textit{$\xi$}$_{VG\mathrm{\bot}}$ would remain finite with
nearly a constant value. This applies to Nb/Cu superlattices in the coupled
regime. Following Yamasaki \textit{et al.}$^{16}$ and Zefrioui \textit{et
al.}$^{17}$ we can estimate an upper limit of \textit{$\xi$}$_{VG\mathrm{\bot
}}$ from I-V characteristics. Isotherms above $T_{g}$ show up Ohmic behavior
at low current level, but they becomes non-linear above $I_{nl}$. At this
current level, the work done by the Lorentz force to create vortex excitations
equals the thermal energy ,$^{4}$ $J_{nl}\phi_{0}\xi_{VG||}\xi_{VG\bot}%
=k_{B}T$. We used the isotherm at the highest temperature within the critical
region, in particular the one at T=4.038 K shown in Fig. 3 (a), for which
$J_{nl}\approx$125 Acm$^{\mathrm{-}\mathrm{2}}$. The in-plane correlation
length \textit{$\xi$}$_{VG\mathrm{|}\mathrm{|}}$ should be larger than the
mean inter-vortex distance $a_{0}=\left(  {{{\phi_{0}}%
\mathord{\left/ {\vphantom {{\phi _{0}}  {\mu
_{0} H}}} \right. \kern-\nulldelimiterspace} {\mu _{0} H}}}\right)
^{{{1}\mathord{\left/ {\vphantom {{1} {2}}} \right. \kern-\nulldelimiterspace}
{2}}}\approx130\;nm$ for $\mu_{0}H=0.11\;T$. In particular we assumed
\textit{$\xi$}$_{VG\mathrm{|}\mathrm{|}}$%
%TCIMACRO{\TEXTsymbol{>}}%
%BeginExpansion
$>$%
%EndExpansion
2$a_{0}$, and thus we obtained \textit{$\xi$}$_{VG\mathrm{\bot}}$%
%TCIMACRO{\TEXTsymbol{<}}%
%BeginExpansion
$<$%
%EndExpansion
90 nm. That is, the correlation length along vortex line is always shorter
than sample thickness. However, it may be longer than other relevant
characteristic lengths, as the superlattice modulation length $\Lambda$=40 nm
and the superconducting coherence length \textit{$\xi$}$_{S\mathrm{\bot}}$%
($T$)$\approx$ 35 nm. We have estimated the vortex length $l$ in the regime
where this superlattice is decoupled, in which we observed a \textit{pure} 2D
VG transition, with the work done by Lorentz \ force $J_{nl}\phi_{0}\xi
_{VG||}l=k_{B}T$.$^{19}$ Taking the isotherm at $T$=2.150 K [Fig. 3 (a)],
$J_{nl}\approx$80 Acm$^{\mathrm{-}\mathrm{2}}$ , and \textit{$\xi$%
}$_{VG\mathrm{|}\mathrm{|}}$%
%TCIMACRO{\TEXTsymbol{>}}%
%BeginExpansion
$>$%
%EndExpansion
2$a_{0}=$150 nm for $\mu_{0}H=0.4\;T$, we get $l<$30 nm. Therefore, the vortex
length $l$ is shorter than superlattice modulation length $\Lambda$=40 nm, and
cannot be much longer than the coherence length at this $T$, \textit{$\xi$%
}$_{S\mathrm{\bot}}\approx$ 20 nm. A picture emerges from those estimations,
in which coherence length \textit{$\xi$}$_{S\mathrm{\bot}}$and sample
thickness $d$ are the relevant length scales to which correlation length along
vortex line \textit{$\xi$}$_{VG\bot}$ has to be compared. The \textit{quasi}%
-2D character of the glass transition in the coupled regime develops since
\textit{$\xi$}$_{VG\bot}$, longer that \textit{$\xi$}$_{S\mathrm{\bot
}\mathrm{,}}$ is shorter than sample thickness $d$. At lower temperatures, Nb
layers are decoupled by Cu ones. Therefore the vortex length $l$ is limited
below modulation length $\Lambda$. Because of this, the vortex length $l$ and
coherence length \textit{$\xi$}$_{S\mathrm{\bot}}$.are similar This yields a
\textit{pure} 2D VG transition. This is similar to what is observed in highly
anisotropic HTCS, for which \textit{pure} 2D VG transitions have been observed
when vortex length is limited to superconducting coherence length.$^{17,20,21}%
$

In summary, we have shown strong evidence of the VG transition in the mixed
state of low-temperature superconducting Nb/Cu superlattices. The HTCS
\ analysis applies directly to this system, and the same universality has been
observed, in spite of the very small thermal fluctuations of LTCS in
comparison with HTCS. Besides, a dimensional crossover from a \textit{quasi}%
-2D into a \textit{pure} 2D VG transition has been observed, which is governed
by the ratio of the VG correlation length $\xi_{\mathrm{V}\mathrm{G}}$ to the
superconducting coherence length $\xi_{\mathrm{S}}$.

We acknowledge financial support from Spanish Ministerio de Educaci\'{o}n y
Ciencia under grants MAT2002-04543 and MAT2002- 12385-E, and ''Ram\'{o}n
Areces'' Foundation. We would like to acknowledge Z. Sefrioui and J.
Santamaria for discussions.\newpage

\bigskip REFERENCES

$^{1}$F. Bouquet \textit{et al.} , Nature \textbf{411}, 449 (2001).

$^{2}$G. Blatter et al., Rev. Mod. Phys. \textbf{64}, 1125 (1994).

$^{3}$M.P.A. Fisher, Phys. Rev. Lett. \textbf{62,} 1415 (1989).

$^{4}$D. S. Fisher, M.P.A. Fisher and D.A. Huse, Phys. Rev. B \textbf{43} 130 (1991).

$^{5}$J.W. Lynn \textit{et al}, Phys. Rev. Lett. \textbf{72}, 3413 (1994).

$^{6}$P.L. Gammel \textit{et al.,} Phys. Rev Lett. \textbf{80}, 833 (1998).

$^{7}$X.S. Ling \textit{et al.}, Phys. Rev. Lett. \textbf{86}, 712 (2001)

$^{8}$M.F. Schmidt, N.E. Israeloff and A.M. Goldman, Phys. Rev. Lett
\textbf{70}, 2162 (1993).

$^{9}$Y. Ando, H. Kubota and S. Tanaka, Phys. Rev. B. \textbf{48}, 7716 (1993).

$^{10}$C.S.L. Chun, G.-G. Zheng, J.L. Vicent, I. K. Schuller, Phys. Rev. B
\textbf{29,} 4915 (1984).

$^{11}$D.R. Strachan \textit{et al.}, Phys. Rev. Lett. \textbf{87}, 067007 (2001).

$^{12}$I.K. Schuller, Phys. Rev. Lett. \textbf{44,} 1597 (1980); J.-P. Locquet
\textit{et al.}, Phys. Rev. B. \textbf{38}, 3572 (1988); J.-P. Locquet
\textit{et al}. Phys. Rev. B. \textbf{39}, 13338 (1989).

$^{13}$E.E. Fullerton \textit{et al.}, Phys. Rev. B \textbf{45}, 9292 (1992).

$^{14}$P. J. M. W\"{o}ltgens \textit{et al.,} Phys. Rev. B 52, 4536 (1995).

$^{15}$P. Voss-deHaan, G. Jakob, and H. Adrian, Phys. Rev. B \textbf{60}, 12
443 (1999).

$^{16}$H. Yamasaki \textit{et al.,} Phys. Rev. B \textbf{50}, 12959 (1994).

$^{17}$Z. Sefrioui \textit{et al}., Phys. Rev. B \textbf{60,} 15423 (1999); E.
M. Gonzalez \textit{et al.,} J. Low Temp. Phys. \textbf{117,} 675 (1999).

$^{18}$K. Moloni \textit{et al.}, Phys. Rev. B \textbf{56}, 14784 (1997).

$^{19}$C. Dekker \textit{et al.}, Phys. Rev. Lett \textbf{69}, 2717 (1992).

$^{20}$Hai-hu Wen \textit{et al}, Phys. Rev. Lett. \textbf{80}, 3859 (1998).

$^{21}$Z. Sefrioui \textit{et al.}, Europhys. Lett. \textbf{49}, 679 (1999).

.

\newpage%
%TCIMACRO{\FRAME{ftbpFU}{5.6178in}{6.0226in}{0pt}{\Qcb{Superconducting
%coherence lengths of sample Cu$_{\mathrm{2}\mathrm{7}\mathrm{n}\mathrm{m}}%
%$[Nb$_{\mathrm{1}\mathrm{3}\mathrm{n}\mathrm{m}}$/Cu$_{\mathrm{2}%
%\mathrm{7}\mathrm{n}\mathrm{m}}$]$_{\mathrm{1}\mathrm{0}}$as a function on
%temperature, both parallel \textit{$\xi$}$_{S||}(T)$ and perpendicular
%\textit{$\xi$}$_{S\bot}(T)$ to Nb/Cu layers. Inset: Measured parallel (black
%circles) and perpendicular (open circles) critical fields H$_{\mathrm{c}%
%\mathrm{2}}$. Solid lines are linear fits $H_{c2\bot}$\textit{(T)$\propto
%$(1-T/T}$_{c})$, while the dashed one is the best fit to $H_{c2||}%
%$\textit{(T)$\propto$(1-T/T}$_{c})^{1/2}.$}}{}{fig1r.ps}%
%width 5.6178in;  height 6.0226in;  depth 0pt;  original-width 8.0004in;
%original-height 10.6666in;  cropleft "0";  croptop "1";  cropright "1.2434";
%cropbottom "0";  filename 'Fig1r.ps';file-properties "XNPEU";}} }%
%BeginExpansion
\begin{figure}
[ptb]
\begin{center}
\includegraphics[
trim=0.000000in 0.000000in -1.947297in 0.000000in,
natheight=10.666600in,
natwidth=8.000400in,
height=6.0226in,
width=5.6178in
]%
{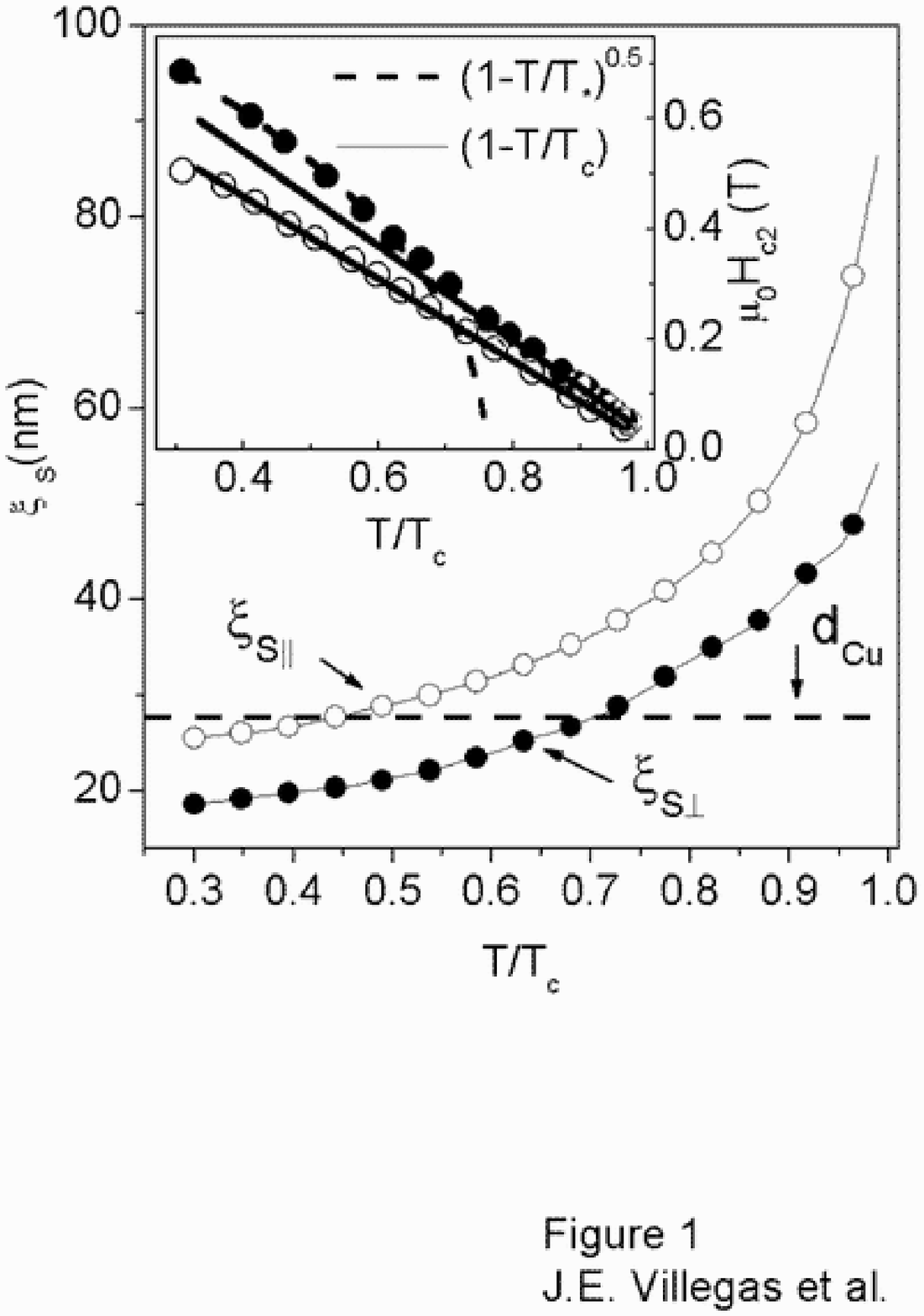}%
\caption{Superconducting coherence lengths of sample Cu$_{\mathrm{2}%
\mathrm{7}\mathrm{n}\mathrm{m}}$[Nb$_{\mathrm{1}\mathrm{3}\mathrm{n}%
\mathrm{m}}$/Cu$_{\mathrm{2}\mathrm{7}\mathrm{n}\mathrm{m}}$]$_{\mathrm{1}%
\mathrm{0}}$as a function on temperature, both parallel \textit{$\xi$}%
$_{S||}(T)$ and perpendicular \textit{$\xi$}$_{S\bot}(T)$ to Nb/Cu layers.
Inset: Measured parallel (black circles) and perpendicular (open circles)
critical fields H$_{\mathrm{c}\mathrm{2}}$. Solid lines are linear fits
$H_{c2\bot}$\textit{(T)$\propto$(1-T/T}$_{c})$, while the dashed one is the
best fit to $H_{c2||}$\textit{(T)$\propto$(1-T/T}$_{c})^{1/2}.$}%
\end{center}
\end{figure}
%EndExpansion

\newpage%
%TCIMACRO{\FRAME{fbpFU}{5.6204in}{6.0226in}{0pt}{\Qcb{(a) I-V isotherms for
%sample Cu$_{\mathrm{2}\mathrm{7}\mathrm{n}\mathrm{m}}$[Nb$_{\mathrm{1}%
%\mathrm{3}\mathrm{n}\mathrm{m}}$/Cu$_{\mathrm{2}\mathrm{7}\mathrm{n}%
%\mathrm{m}}$]$_{\mathrm{1}\mathrm{0}}$ in applied field $\mu_{\mathrm{0}}%
%$H=0.11 T at temperatures (from left to right) 4.060 K\TEXTsymbol{>}%
%T\TEXTsymbol{>}3.865 K, separated $\sim$5-15 mK. The dashed line separates
%isotherms above and below T$_{\mathrm{g}}$=3.994 K. Vertical and horizontal
%dotted lines delimit the experimental window used for scaling. (b) Scaling of
%the above isotherms as explained in the text . Inset: Derivatives of the
%log(I)-log(V) isotherms at temperatures (from bottom to top) 4.060
%K\TEXTsymbol{>}T\TEXTsymbol{>}3.908 K. Black dots are within the experimental
%window used for scaling. The arrow separates derivatives of isotherms just
%below and above \textit{T}$_{g}.$}}{}{fig2r.ps}%
%width 5.6204in;  height 6.0226in;  depth 0pt;  original-width 8.1664in;
%original-height 10.6666in;  cropleft "0";  croptop "1";  cropright "1.2187";
%cropbottom "0";  filename 'FIg2r.ps';file-properties "XNPEU";}} }%
%BeginExpansion
\begin{figure}
[pb]
\begin{center}
\includegraphics[
trim=0.000000in 0.000000in -1.785992in 0.000000in,
natheight=10.666600in,
natwidth=8.166400in,
height=6.0226in,
width=5.6204in
]%
{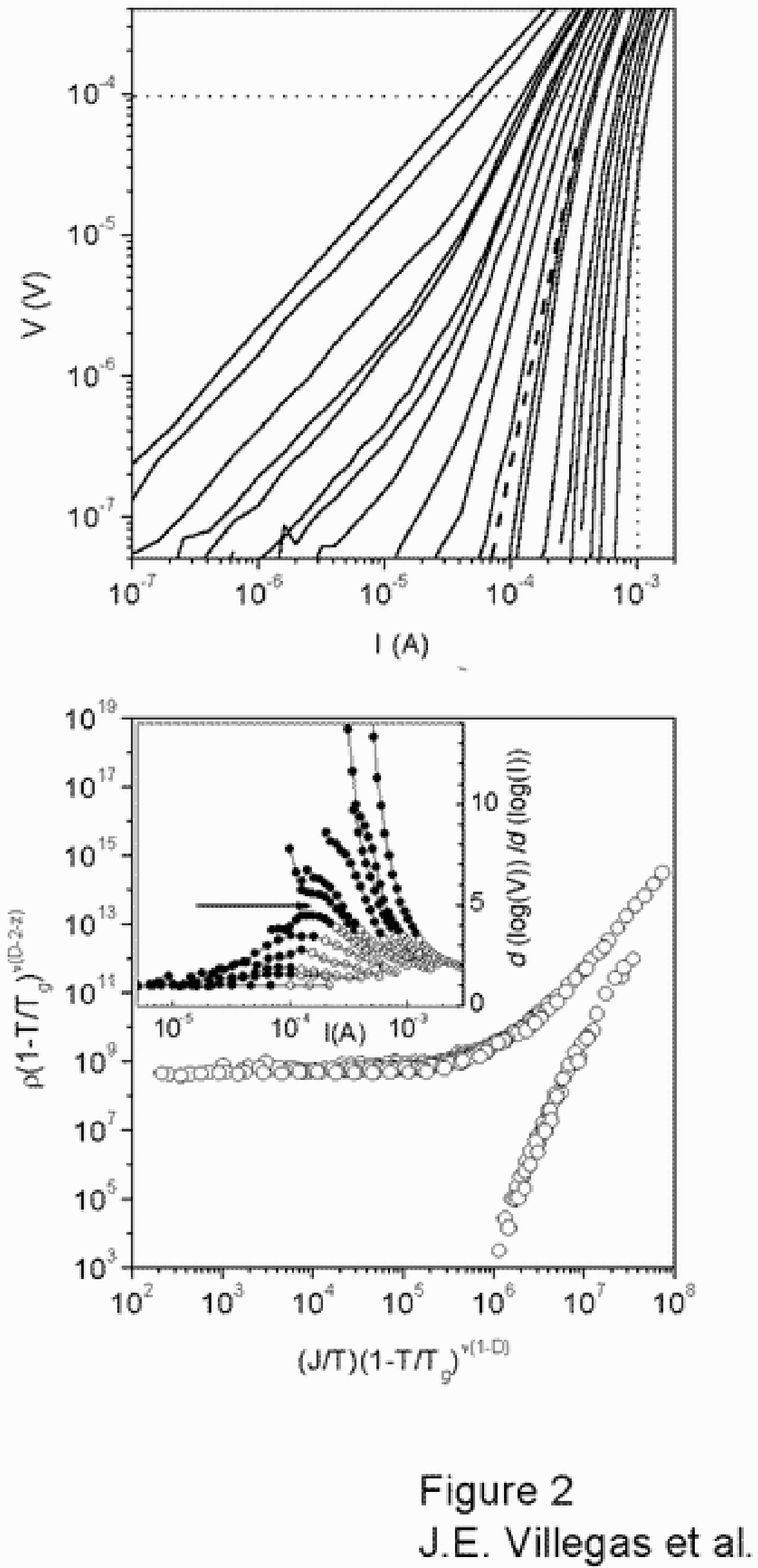}%
\caption{(a) I-V isotherms for sample Cu$_{\mathrm{2}\mathrm{7}\mathrm{n}%
\mathrm{m}}$[Nb$_{\mathrm{1}\mathrm{3}\mathrm{n}\mathrm{m}}$/Cu$_{\mathrm{2}%
\mathrm{7}\mathrm{n}\mathrm{m}}$]$_{\mathrm{1}\mathrm{0}}$ in applied field
$\mu_{\mathrm{0}}$H=0.11 T at temperatures (from left to right) 4.060 K$>$%
T$>$3.865 K, separated $\sim$5-15 mK. The dashed line separates isotherms
above and below T$_{\mathrm{g}}$=3.994 K. Vertical and horizontal dotted lines
delimit the experimental window used for scaling. (b) Scaling of the above
isotherms as explained in the text . Inset: Derivatives of the log(I)-log(V)
isotherms at temperatures (from bottom to top) 4.060 K$>$T$>$3.908 K. Black
dots are within the experimental window used for scaling. The arrow separates
derivatives of isotherms just below and above \textit{T}$_{g}.$}%
\end{center}
\end{figure}
%EndExpansion

\newpage%

%TCIMACRO{\FRAME{ftbpFU}{5.7683in}{6.0226in}{0pt}{\Qcb{(a) Scaling of the
%isotherms shown in the inset with a \textit{pure} 2D VG model. Inset: I-V
%isotherms for sample Cu$_{\mathrm{2}\mathrm{7}\mathrm{n}\mathrm{m}}%
%$[Nb$_{\mathrm{1}\mathrm{3}\mathrm{n}\mathrm{m}}$/Cu$_{\mathrm{2}%
%\mathrm{7}\mathrm{n}\mathrm{m}}$]$_{\mathrm{1}\mathrm{0}}$ in applied field
%$\mu_{\mathrm{0}}$H=0.4 T at temperatures (from left to right) 2.160
%K\TEXTsymbol{>}T\TEXTsymbol{>}1.780 K, separated $\sim$5-15 mK.. Horizontal
%dotted line delimit the experimental window used for scaling. (b) Derivatives
%of the log(I)-log(V) isotherms at temperatures (from bottom to top) 2.160
%K\TEXTsymbol{>}T\TEXTsymbol{>}1.856 K. Black dots are within the experimental
%window used for scaling. The arrow marks derivatives of the isotherm at the
%lowest temperature showing decreasing slope in the low current limit}}%
%{}{fig3r.ps}{\special{ language "Scientific Word";  type "GRAPHIC";
%width 5.7683in;  height 6.0226in;  depth 0pt;  original-width 8.1664in;
%original-height 10.6666in;  cropleft "0";  croptop "1";  cropright "1.2507";
%cropbottom "0";  filename 'Fig3r.ps';file-properties "XNPEU";}} }%
%BeginExpansion
\begin{figure}
[ptb]
\begin{center}
\includegraphics[
trim=0.000000in 0.000000in -2.047317in 0.000000in,
natheight=10.666600in,
natwidth=8.166400in,
height=6.0226in,
width=5.7683in
]%
{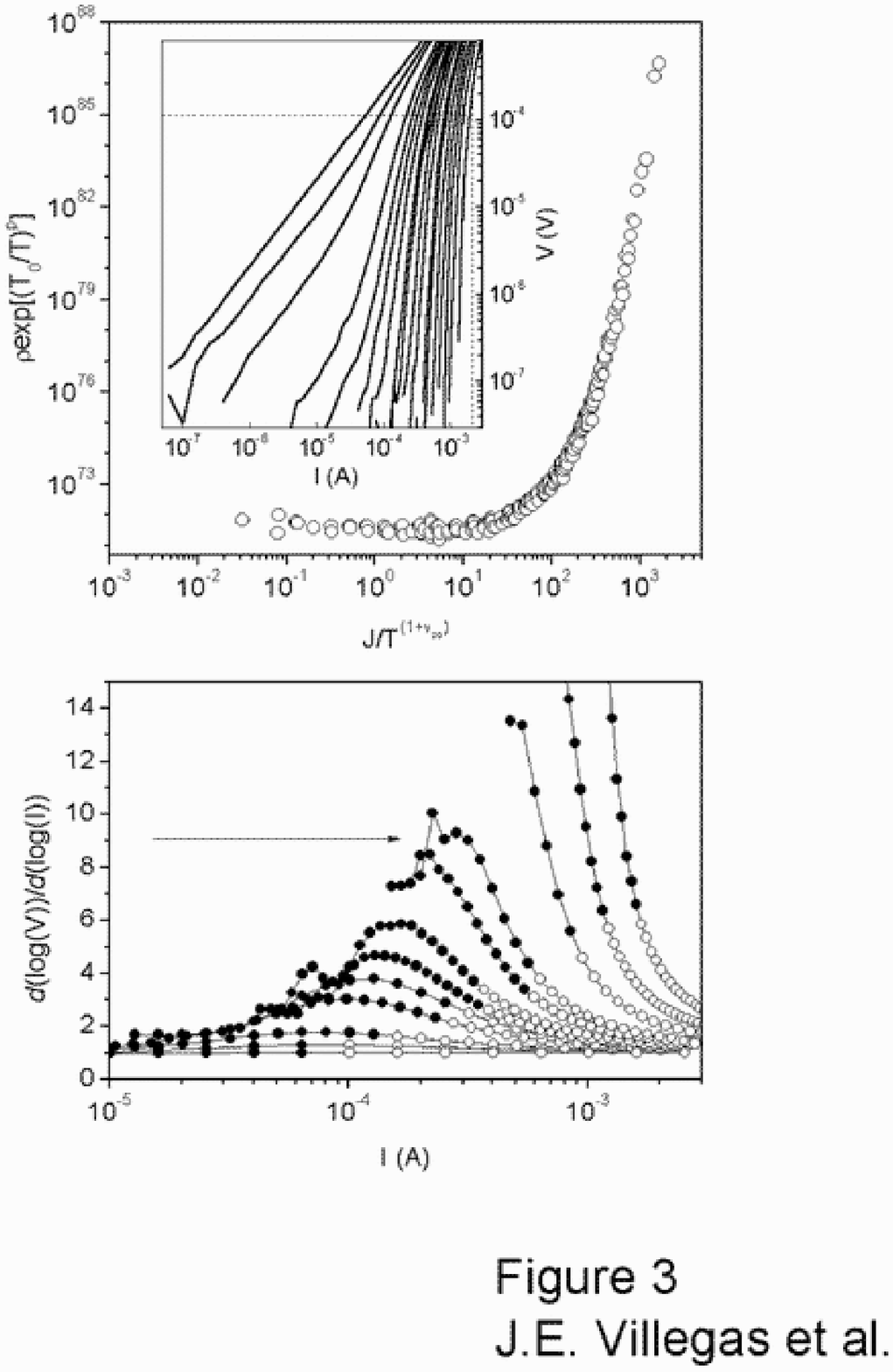}%
\caption{(a) Scaling of the isotherms shown in the inset with a \textit{pure}
2D VG model. Inset: I-V isotherms for sample Cu$_{\mathrm{2}\mathrm{7}%
\mathrm{n}\mathrm{m}}$[Nb$_{\mathrm{1}\mathrm{3}\mathrm{n}\mathrm{m}}%
$/Cu$_{\mathrm{2}\mathrm{7}\mathrm{n}\mathrm{m}}$]$_{\mathrm{1}\mathrm{0}}$ in
applied field $\mu_{\mathrm{0}}$H=0.4 T at temperatures (from left to right)
2.160 K$>$T$>$1.780 K, separated $\sim$5-15 mK.. Horizontal dotted line
delimit the experimental window used for scaling. (b) Derivatives of the
log(I)-log(V) isotherms at temperatures (from bottom to top) 2.160 K$>$%
T$>$1.856 K. Black dots are within the experimental window used for scaling.
The arrow marks derivatives of the isotherm at the lowest temperature showing
decreasing slope in the low current limit}%
\end{center}
\end{figure}
%EndExpansion

\end{document}